# Dynamic routing based on call quality.

Routage Dynamique basé sur la qualité des appels.

**Oussama Hammami, Christian Lathion, Emin Gabrielyan**

*Switzernet Sarl, Parc Scientifique (PSE) de l'Ecole Polytechnique Fédérale de Lausanne (EPFL)*
PSE-A, CH-1015, Lausanne Suisse
oussama.hammam@switzernet.com, christian.lathion@switzernet.com, emin.gabrielyan@switzernet.com

ABSTRACT
The telephony over IP (ToIP) is becoming a new trend in technology widely used nowadays in almost all business sectors. Its concepts rely on transiting the telephone communications through the IP network. Today, this technology is deployed increasingly what the cause of emergence of companies is offering this service as Switzernet.
For several highly demanded destinations, recently fake vendors appeared in the market offering voice termination but providing only false answer supervision. The answered signal is returned immediately and calls are being charged without being connected. Different techniques are used to keep the calling party on the line. One of these techniques is to play a record of a ring back tone (while the call is already being charged). Another, more sophisticated technique is to play a human voice randomly picked up from a set of records containing contents similar to: hello, hello, I cannot hear you
 Apart the fact that the fallaciously established calls are charged at rates of real calls, such malicious routes seriously handicap the switching process. The system does not detect a failure on signaling level and is unable to attempt the call via backup routes, the call technically being already connected. Once the call flow falls into such trap, the calls will continue being routed via the fraudulent route until a manual intervention.

RESUME
La téléphonie sur IP (ToIP ou VoIP) est une technologie qui s'impose progressivement dans tous les secteurs, elle consiste à faire transiter les communications téléphoniques par le réseau IP.
Aujourd'hui, cette technologie est de plus en plus déployée ce qui est la cause de l'apparition des entreprises comme Switzernet qui offre ce service
Toutefois, les aspects techniques sous-jacents à cette nouvelle technologie ne sont pas toujours bien maîtrisés. Les problèmes dus au NAT, les pare-feux, la sécurité, mauvaise signalisation côté fournisseur, etc. sont des problèmes qui restent encore à dominer.
Afin de résoudre le problème de mauvaise signalisation côté fournisseur on a décidé de router dynamiquement les appels vers le vendeur qui offre une meilleur qualité au lieu du routage statique basé sur des préférences prédéfini dans le serveur de facturation.
Cette solution repose sur l'ajout un autre serveur de routage 'Kamailio'.

*MOTS-CLES: SIP, ToIP, VoIP, Routage dynamique, ACD, Kamailio, OpenSer, Billing, Portasip.*

# Table des matières

# Introduction générale

La téléphonie sur IP constitue actuellement une des plus importantes évolutions dans le domaine des Télécommunications. Il y a quelques années, la transmission de la voix sur le réseau téléphonique classique ou RTC constituait l'exclusivité des télécommunications.

Aujourd'hui, les données ont changé. La transmission de la voix via les réseaux IP constitue une nouvelle évolution majeure comparable à la précédente. Au delà de la nouveauté technique, la possibilité de fusion des réseaux IP et téléphoniques entraîne non seulement une diminution de la logistique nécessaire à la gestion des deux réseaux, mais aussi une baisse importante des coûts de communication ainsi que la possibilité de mise en place de nouveaux services utilisant simultanément la voix et les données.

Après l'étude du problème de la mauvaise qualité de la voix pour les appels à destination de l'Arménie – Yerevan on a constaté que ce problème est dû aux faux signaux envoyés par les fournisseurs des appels entrants/sortants.

Généralement lorsqu'un vendeur tombe en panne il répond par un message vocal d'erreur mais il n'envoie pas le signal d'erreur de type 4xx ,5xx où 6xx (voir Fig.1) pour que l'appel sera routé vers le deuxième vendeur, l'appel sera facturé et considéré comme appel réussi, la seule différence avec un vrai appel c'est que sa durée n'est pas longue (quelque secondes après l'utilisateur raccroche) ce qui fait chuter forcément la valeur de L'ACD d'où on a décidé de router les appels en se basant sur ces valeurs.

En effet le serveur de facturation (Billing) continue toujours de faire le routage selon les préférences prédéfinies mais au lieu d'envoyer les appels vers les vrais vendeurs on les envoie vers l'interface réseaux qui leurs correspond dans le serveur Kamailio et ce serveur va décider s'il passe l'appel à ce vendeur ou non en tenant compte toujours de la qualité offerte par ce dernier pendant l'intervalle précédent.

Cette solution se résume en 2 chapitres principaux.
Le 1ème chapitre donnera un aperçu global sur les protocoles associés à la téléphonie sur IP.
Le 2ème chapitre présentera et expliquera notre solution.

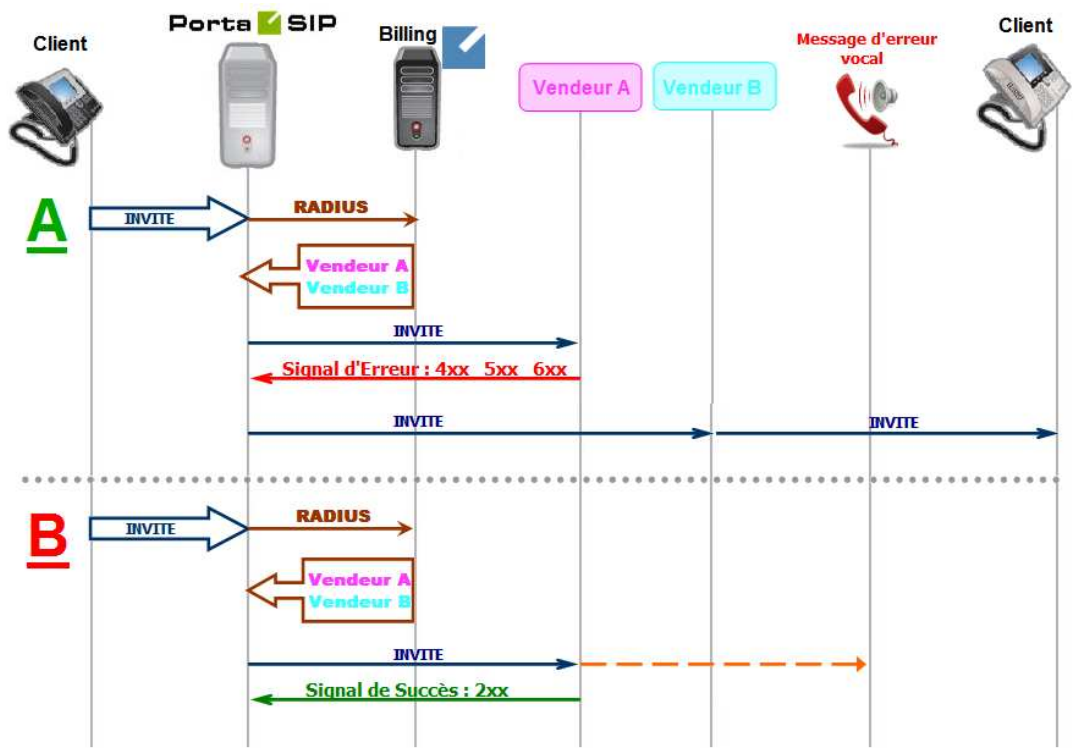

**Fig.1** Problème de Signalisation.

# CHAPITRE I : Etat de l'art des protocoles VoIP

## *1. Introduction*

La téléphonie sur IP est une technologie de communication vocale en pleine émergence.
Elle fait partie d'un tournant dans le monde de la communication. En effet, la convergence du triple Play (voix, données et vidéo) fait partie des enjeux principaux des acteurs de la télécommunication aujourd'hui.
La VoIP possède actuellement une véritable opportunité économique pour les entreprises telles que la diminution du coût en infrastructure, de la facture de téléphone.
La téléphonie sur IP est basée sur des standards ouverts : elle permet donc l'interaction avec les équipements téléphoniques standards. Toutefois, les aspects techniques sous-jacents à cette nouvelle technologie ne sont pas toujours bien maîtrisés.
Les problèmes dus au NAT, les pare-feux, la sécurité, etc. sont des problèmes qui restent encore à dominer.

## *2. Protocoles liés à la VoIP*

La voix sur IP ou VoIP (Voice over IP) est un service de téléphonie qui transporte les flux voix des communications téléphoniques sur un réseau IP.
Afin de rendre possibles les communications VoIP, les solutions proposées dopent la couche IP par des mécanismes supplémentaires nécessaires pour apporter la QOS (qualité de service) nécessaire au flux voix de types temps réel, en plus de l'intelligence nécessaire à l'exécution de services.
A cet effet, il existe deux types de protocoles principaux utilisés dans la VoIP :
- Protocoles de signalisation
- Protocoles de transport.

## 2.1 Signalisation

La signalisation correspond à la gestion des sessions de communication (ouverture, fermeture, etc.). Le protocole de signalisation permet de véhiculer un certain nombre d'informations notamment:
- Le type de demande (enregistrement d'un utilisateur, invitation à une session multimédia, annulation d'un appel, réponse à une requête, etc.).
- Le destinataire d'un appel.
- L'émetteur.
- Le chemin suivi par le message.

Plusieurs normes et protocoles ont été développés pour la signalisation VoIP, quelques uns sont propriétaires et d'autres sont des standards. Ainsi, les principales propositions disponibles pour l'établissement de connexions en VoIP sont :
- SIP (Session Initiation Protocol) qui est un standard IETF (Internet Engineering Task Force) décrit dans le RFC 3261.
- H323 englobe un ensemble de protocoles de communication développés par l'UIT-T
(Union Internationale des Télécommunications – secteur de la normalisation des
Télécommunications).
- MGCP (Media Gateway Control Protocol) standardisé par l'IETF (RFC 3435).
- SCCP (Skinny Client Control Protocol) est un protocole propriétaire CISCO.

Aujourd'hui le plus répandu d'entre eux est le SIP, ce protocole est largement déployé et utilisé au sein de Switzernet.

## SIP (Session Initiation Protocol) [3]

Le protocole SIP (Session Initialisation Protocol) a été initié par le groupe MMUSIC
(Multiparty Multimedia Session Control) [RFC 2543] est désormais repris et maintenu par le groupe SIP de l'IETF [RFC 3261].
SIP est un protocole de signalisation appartenant à la couche application du modèle OSI. Il a été conçu pour l'ouverture, le maintient et la terminaison de sessions de communications interactives entre des utilisateurs. De telles sessions permettent de réaliser de l'audio, de l'enseignement à distance et de la voix (téléphonie) sur IP essentiellement. Pour l'ouverture d'une session, un utilisateur émet une invitation transportant un descripteur de session permettant aux utilisateurs souhaitant communiquer de négocier sur les algorithmes et codecs à utiliser. SIP permet aussi de relier des stations mobiles en transmettant ou redirigeant les requêtes vers la position courante de la station appelée. Enfin, SIP est indépendant du médium utilisé et aussi du protocole de transport des couches basses.

• **Les messages**

**Le format des messages**
Requête d'un client vers un serveur :

```
Ligne de requête
(Méthode, Requête URI, version SIP)

En tête général, ou de requête, ou d'entité

CRLF (permet de spécifier la fin du champ d'en-têtes, et le
début du corps du message)

Corps du message
```

Requête d'un serveur vers un client

```
Ligne d'état
(version SIP, code d'état, Reason Phrases)

En tête général, ou de réponse, ou d'entité

CRLF (permet de spécifier la fin du champ d'en-têtes, et le
début du corps du message)

Corps du message
```

**Méthode:** ACK ou INVITE ou BYE ou REGISTER…
**Code d'état:** entier codé sur 3 bits indiquant un résultat à l'issue de la réception d'une requête.
**Reason phase**: pour le refus ou de l'acceptation d'une requête.
**En-têtes général**: Accept (permet d'indiquer les types de média qui seront acceptés dans la réponse) ou Accept-Encoding ou Call-ID (identifie une invitation précise)…
**En-têtes de requête:** Priority ou Route…
**En-têtes d'entité**: Content-Encoding ou Content-Length (taille du corps du message envoyé) ou Content-Type (indique les types de média utilisés)
**En-têtes de réponse**: Allow ou Server…
**Corps du message**: Pour les requêtes, un corps est ajouté ou non selon la méthode utilisée. Pour les réponses, le corps du message est obligatoire.

**Les requêtes**
Les échanges entre un terminal appelant et un terminal appelé se font par l'intermédiaire de requêtes.
Les 4 requêtes de base sont les suivantes :

| Nom de la requête | Description |
|---|---|
| INVITE | demande d'ouverture de session |
| ACK | code de confirmation |
| CANCEL | code d'annulation d'INVITE |
| BYE | terminaison de session |

**Les réponses**

Pour ce qui est des codes de réponse, la similarité avec le HTTP ne fait aucun doute :

| Code | Représentation |
|---|---|
| 100 | trying (tentative) |
| 200 | OK |
| 301/302 | moved permanently/temporarily |
| 404 | not found (non trouvé) |

Cependant, d'autres codes étaient nécessaire, notamment pour gérer des aspects plus techniques de la téléphonie, il s'agit de codes ayant une forme supérieure ou égale à x80 :

| Code | Représentation |
|---|---|
| 180 | ringing (sonnerie) |
| 181 | call is being forwarded (renvoi d'appel) |
| 182 | queued (en attente) |
| 183 | session progress (session en cours) |
| 480 | Temporarily Unavailable |
| 481 | Call/Transaction Does Not Exist |
| 482 | Loop Detected |
| 483 | Too Many Hops |
| 484 | Address Incomplete |
| 485 | Ambiguous |
| 486 | Busy Here (occupé) |

Autrement, on peut les regrouper comme suite :

| Code | Représentation |
|---|---|
| 1xx | Information. la requête à été reçue et est en cours de traitement. (Exemple: "180 Ringing"...) |
| 2xx | Succès. La requête a été traitée correctement. (Exemple: "200 OK") |
| 3xx | Réacheminement. Indique qu'une autre intervention est nécessaire pour effectuer l'appel. |
| 4xx | Erreur du client. Le message comporte une erreur et le serveur l'a rejeté. (Exemple: structure de message erronée) |
| 5xx | Erreur du serveur. le serveur n'a pas réussi à traiter la requête. (Exemple: ressource en panne) |
| 6xx | Echec général. La requête ne peut être traitée sur aucun serveur. (Exemple: aucune ressource disponible à l'échelle du réseau) |

La principale différence avec le HTTP est que la notion client-serveur n'existe pas dans l'optique d'une transmission. Un agent SIP (UA, User Agent) pourra envoyer et recevoir des requêtes.

**Exemple**

| Requête INVITE | Réponse à la requête INVITE |
|---|---|
| ```
RECEIVED message from 212.147.8.99:64401:
    INVITE
sip:0215661380@sip8.youroute.net SIP/2.0
    Via: SIP/2.0/UDP
192.168.1.130:65062;branch=z9hG4bK-d8754z-
ff15203e36356c5e-1---d8754z-;rport
    Max-Forwards: 70
``` | ```
SENDING message to 212.147.8.99:64401:
    SIP/2.0 200 OK
    Via: SIP/2.0/UDP
192.168.1.130:65062;received=212.147.8.
99;branch=z9hG4bK-d8754z-
ff15203e36356c5e-1---d8754z-
;rport=64401
``` |



```
      Contact:                                      Record-Route:
<sip:41215040306@212.147.8.99:64401>          <sip:82.103.128.3;ftag=3c528b47;lr>
      To:                                           From: 41215040306
"0215661380"<sip:0215661380@sip8.youroute.    <sip:41215040306@sip8.youroute.net>;tag
net>                                          =3c528b47
      From:                                         To: 0215661380
"41215040306"<sip:41215040306@sip8.yourout    <sip:0215661380@sip8.youroute.net>;
e.net>;tag=3c528b47                           tag=ed73aa90517ebb7cc6389f928119a85a
      Call-ID:                                      Call-ID:
NzQyZmRjMWNjMzEwNWE3ZmRiNmM5NmVjNzI5M2JmYm    NzQyZmRjMWNjMzEwNWE3ZmRiNmM5NmVjNzI5M2J
Y.                                            mYmY.
      CSeq: 2 INVITE                                CSeq: 2 INVITE
      Allow: INVITE, ACK, CANCEL, OPTIONS,          Server: Sippy
BYE, REFER, NOTIFY, MESSAGE, SUBSCRIBE,             Contact: Anonymous
INFO                                          <sip:82.103.128.3:5061>
      Content-Type: application/sdp                 Content-Length: 152
      User-Agent: X-Lite release 1103k              Content-Type: application/sdp
stamp 53621
      Authorization: Digest                         v=0
username="41215040306",realm="82.103.128.3          o=Sippy 145597708 1 IN IP4
",                                            82.103.128.3
nonce="7e40fe2c44b5e7d299b4424a473058944ad          s=-
71dda",                                             t=0 0
uri="sip:0215661380@sip8.youroute.net",             m=audio 35906 RTP/AVP 0
response="3f987e14e4a79d11661847906d9d2fbc          c=IN IP4 82.103.128.3
",algorithm=MD5                                     a=sendrecv
      Content-Length: 265                           a=ptime:10
                                                    a=rtpmap:0 PCMU/8000
      v=0
      o=- 8 2 IN IP4 192.168.1.130
      s=CounterPath X-Lite 3.0
      c=IN IP4 192.168.1.130
      t=0 0
      m=audio 8680 RTP/AVP 107 0 8 101
      a=alt:1 1 : WMrQlZK2 9YFczmZs
192.168.1.130 8680
      a=fmtp:101 0-15
      a=rtpmap:107 BV32/16000
      a=rtpmap:101 telephone-event/8000
      a=sendrecv
```

- *Mode de transmission*

Le protocole SIP est destiné à des réseaux IP et utilise habituellement le port 5060 (TCP/UDP). Il n'est pas en charge du transport de la voix, il ne gère que l'établissement de l'appel et sa signalisation. Le canal voix à proprement parler utilise généralement le protocole RTP (Real-time Transport Protocol).

- *Identification des agents SIP*

Un agent SIP s'identifie auprès de son Registrar ou, à défaut, du proxy SIP en fournissant son adresse IP. Chaque téléphone, qu'il soit physique (hard) ou logiciel (soft) dispose d'un user agent (un peu comme les navigateurs Internet). Le couple adresse IP-User Agent est donc transmis au Registrar qui, en retour, va enregistrer la localisation du dispositif et lui attribuer une URI SIP. L'URI ou Uniform Resource Identifier ressemble à une adresse e-mail : sip:utilisateur@domain.com.

L'utilisation d'URI au lieu d'adresses IP a plusieurs avantages :

- plus simples à mémoriser qu'une adresse IP
- permet de s'affranchir des problèmes d'adressage par l'utilisation d'un serveur DHCP



- gestion d'adresses publiques/privées

Même si, en théorie, il est possible de faire du peer-to-peer et d'appeler directement d'un téléphone à l'autre, l'utilisation d'URI impose de passer par un registrar ou un proxy SIP qui lui-seul connaît l'emplacement du téléphone.

### • *Le Registrar*

Couplé ou intégré au proxy SIP, il gère les requêtes REGISTER et inscrit dans sa base de données les User Agent. Afin de garantir qu'un utilisateur est joignable, la requête REGISTER est renouvelée régulièrement (entre 60 et 3600 secondes selon la configuration et l'architecture du réseau). Ce mécanisme apporte l'assurance qu'une requête envoyée sur un téléphone aboutira. Si un User Agent ne se réenregistre pas dans la période demandée, il sera supprimé de la base de données. Il devient alors possible de diffuser un message d'indisponibilité ou d'effectuer un renvoi vers une messagerie par exemple.

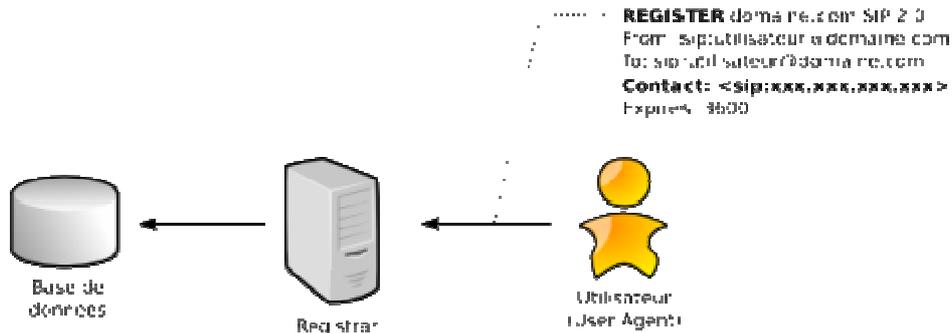

**Fig.2 :** Illustration de l'enregistrement d'un utilisateur avec un temps d'expiration de 3600s

### • *L'appel SIP*

Une fois l'appel négocié, le protocole SIP ne fait pas transiter l'appel en lui-même, il est transmis directement entre les User Agents ou entre l'User Agent et la passerelle de/vers le réseau public dans le cas d'un appel externe.

Un appel SIP transitant par un proxy SIP se déroule de la manière suivante :

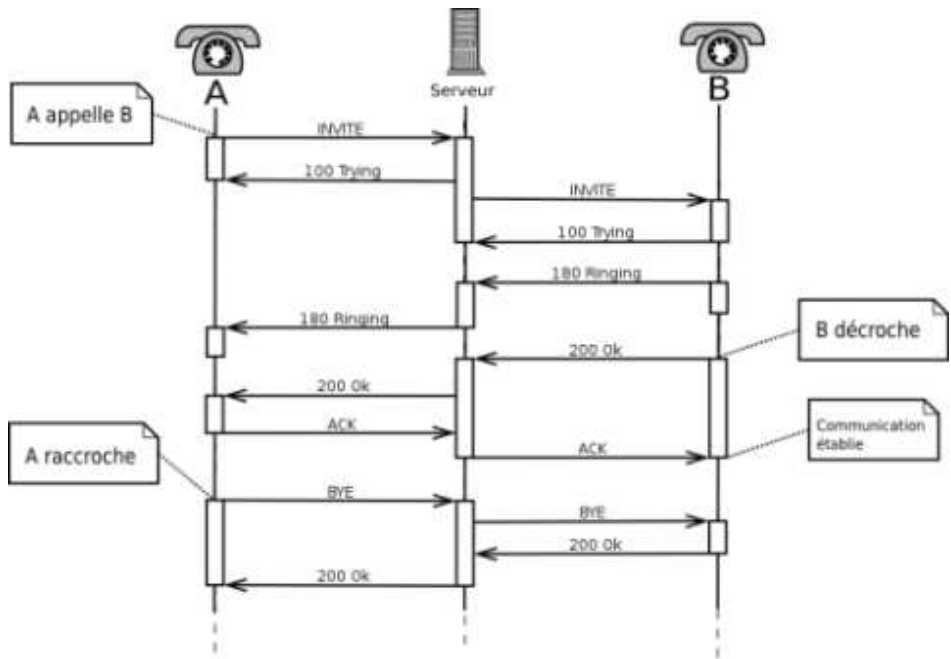

**Fig.3 :** Illustration d'un scénario d'appel SIP



![Fig.4 SIP packets capture from Wireshark showing INVITE, 100 trying, 401 Unauthorized, ACK, 180 Ringing, 200 OK, REGISTER, BYE exchanges between 192.168.1.130 and 82.103.128.3]

**Fig.4:** les paquets SIP (avec Wireshark)

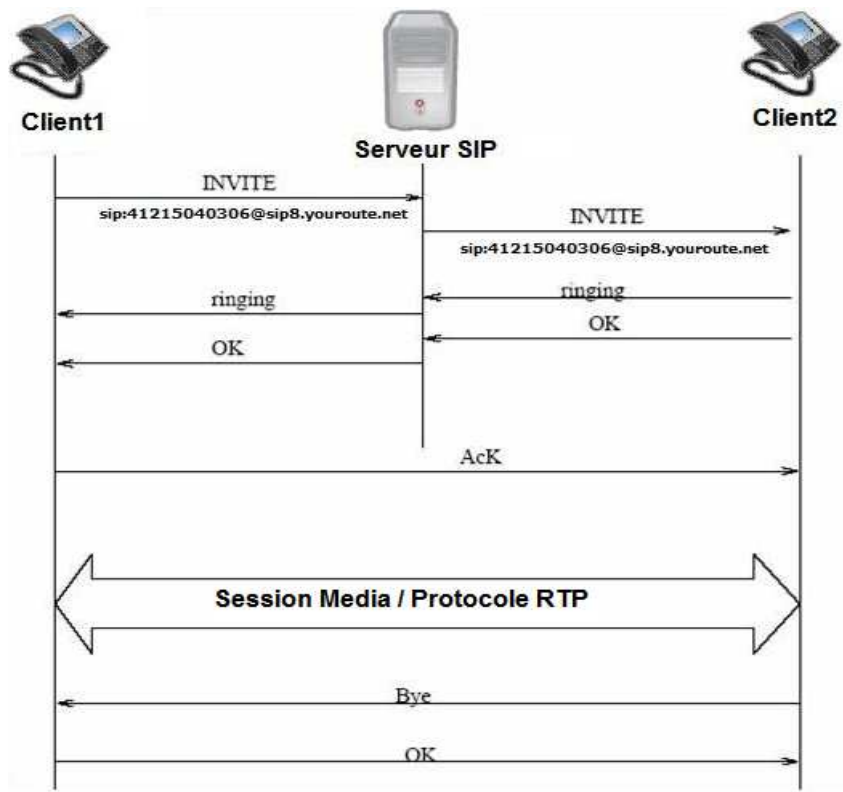

**Fig.5:** Illustration d'un appel transitant par un proxy SIP



## 2.2 Transport

Lors d'une communication VoIP, une fois la phase de signalisation réalisée, la phase de communication est initiée. Dans cette phase, un protocole de transport permet d'acheminer les données voix entre plusieurs utilisateurs vu que la couche TCP propose un transport fiable mais lent, et la couche UDP un transport rapide mais non fiable. La communauté IETF a mis en place un nouveau couple de protocole RTP (Real-Time transport Protocol) et RTCP
(Real-Time Control Protocol) pour apporter la fiabilité à l'UDP tout en exploitant sa rapidité.
RTP et RTCP sont les deux protocoles qui sont principalement utilisés pour le transport de flux média sur le réseau IP. RTP permet de transporter les données entre plusieurs utilisateurs en plus de la gestion temps réelle des sessions. Tandis que, RTCP est utilisé pour transmettre régulièrement des paquets de contrôle, qui contiennent diverses statistiques, ce qui permet de vérifier la qualité de transmission.

## *3. Conclusion*

La téléphonie sur IP est une technologie qui utilise les réseaux informatiques comme support de communication. Les solutions VoIP sont de plus en plus basées sur des standards ouverts. Beaucoup de ces solutions utilisent SIP comme protocole de signalisation VoIP. Les principaux protocoles utilisés pour le transport de la voix sont : RTP et RTCP.



# CHAPITRE II: Design et ingénierie VoIP

## *1. Présentation de l'existant*

### 1.1 Infrastructure Switzernet

L'architecture de Switzernet comprend deux types de serveurs. Les serveurs SIP qui font transiter les appels, et les serveurs de Billing qui gèrent le routage et la facturation :

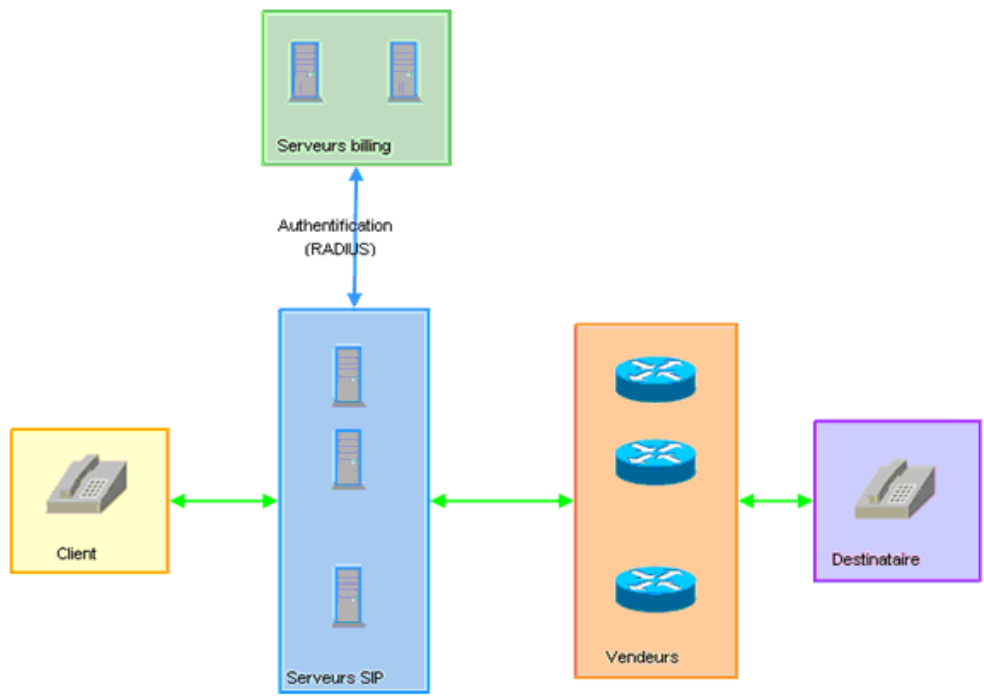

**Fig.6 :** Infrastructure Switzernet

Pour les clients, seul le serveur SIP (sipX.youroute.net) est visible. Pour autoriser l'appel, les serveurs SIP interrogent les serveurs de facturation. Selon les informations de la base de données Billing (balance, client bloqué ou non, etc.), l'appel sera autorisé ou refusé.

La communication entre les serveurs SIP et les serveurs de Billing se fait par le protocole RADIUS. C'est la partie AAA (Authentication, Authorization, Accounting) de RADIUS.

### 1.2 Principe de fonctionnement

En résumé, un appel sortant passe par les étapes suivantes :

  - Le serveur SIP reçoit la demande d'appel (message INVITE) du client
  - <u>Authentication</u> : Il interroge le serveur de Billing pour vérifier le nom d'utilisateur et mot de passe
  - <u>Authorization</u> : Il interroge le serveur de Billing pour savoir si l'appel est autorisé (compte pas bloqué, balance suffisante, etc.)
  - Si l'appel est autorisé, le serveur Billing indique par quel vendeur le faire transiter
  - Le serveur SIP établit l'appel avec le vendeur qui va l'envoyer au destinataire final
  - <u>Accounting</u> : Au moment de la fin d'appel, le serveur SIP indique au serveur de Billing la fin d'appel, la durée, etc.

• *Serveurs SIP [sip_servers]*



Ce sont les **serveurs SIP** qui gèrent les appels. On en utilise plusieurs sur lesquels sont distribués les clients, pour répartir la charge et minimiser l'impact des pannes. Chaque serveur peut avoir plusieurs noms (enregistrements DNS) : un ou plusieurs noms "publics" (ex. **sip1.youroute.net**) et un nom "interne" (ex. **fr1.youroute.net**). Au moment de l'inscription, on assigne un serveur SIP à chaque client (**sip1.youroute.net**, **sip2.youroute.net**, etc.), mais en vérité le client peut s'inscrire sur n'importe quel serveur. D'ailleurs certains des différents noms **sipX.youroute.net** pointent en fait vers le même serveur.

• *Vendeurs*
Les vendeurs sont nos fournisseurs pour les appels entrants/sortants. C'est en général de leur côté que se trouvent les lignes téléphoniques physiques, on leur envoie les appels directement par Internet.

## *2. Travail effectué*

### 2.1 Objectifs
L'objectif principal était de proposer une solution qui permet de :
   - Router dynamiquement les appels à destination de l'Arménie – Yerevan vers le vendeur qui offre une meilleur qualité.
   - Surveiller l'état du service de routage des appels dans le serveur Kamailio.

### 2.2 Mise en place d'une solution de Routage dynamique
Le Billing route les appels vers le vendeur dont la préférence est la plus grande mais ce type de routage n'est pas faible car on envoie les appels vers un vendeur quelque soit son état (chargée, en panne) ce qui provoque parfois des indisponibilités ou problèmes de qualité.
Par contre pour ce nouveau routage, on va tenir compte de l'ACD (Moyenne de durée d'appel) pour un intervalle dynamique ainsi que la préférence pour calculer une probabilité de routage pour chaque vendeur.
Pour la mise en production d'un service de routage dynamique d'appels pour l'Arménie, un serveur Kamailio a été mis en place.

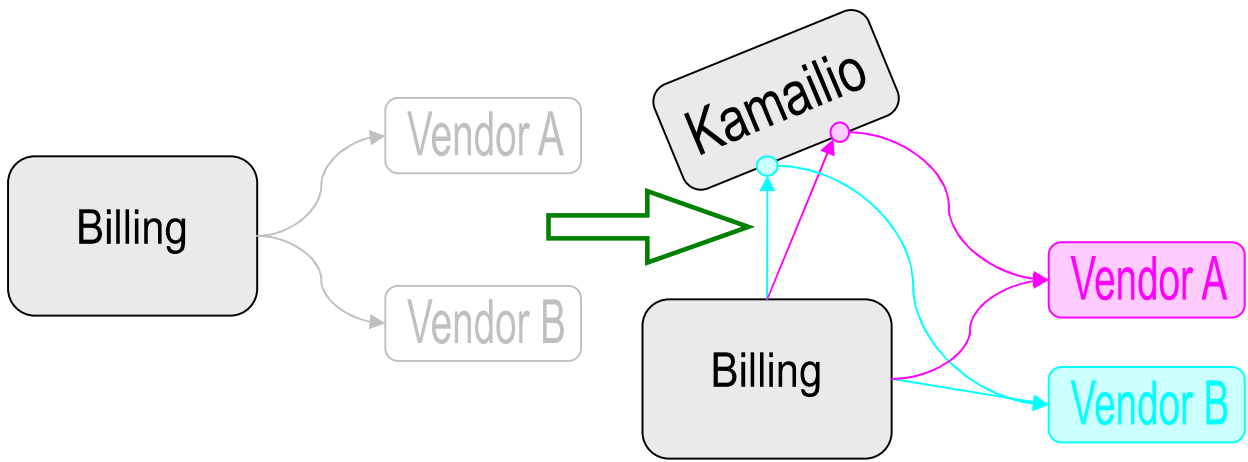

**Fig.7 :** Architecture de la solution mise en place

• *Présentation de Kamailio* **[4]**
Kamailio est un fork de SIP Express Router (SER). Ils présentent donc de très grandes similitudes, et leurs fichiers de configuration sont en grande partie compatibles.

Les fichiers de configuration de Kamailio sont installés par défaut dans le répertoire 'etc/kamailio/'. Le fichier principal est 'kamailio.cfg'. La configuration se fait en écrivant un script qui décrit les actions à effectuer pour chaque paquet reçu. On a donc une très grande flexibilité pour gérer les appels SIP.

Ce fichier de configuration Kamailio (kamailio.cfg) est plus qu'un fichier de configuration typique. Il combine à la fois les paramètres statiques et un environnement de programmation dynamique.

En effet, 'kamailio.cfg' est un programme qui est exécuté pour chaque message reçu par le routeur Open SIP Express (OpenSER nommé Kamailio à partir de la version 1.4.0).



• *Architecture :*
Le schéma suivant illustre le scénario de deux appels transitant par notre serveur de routage (Kamailio).

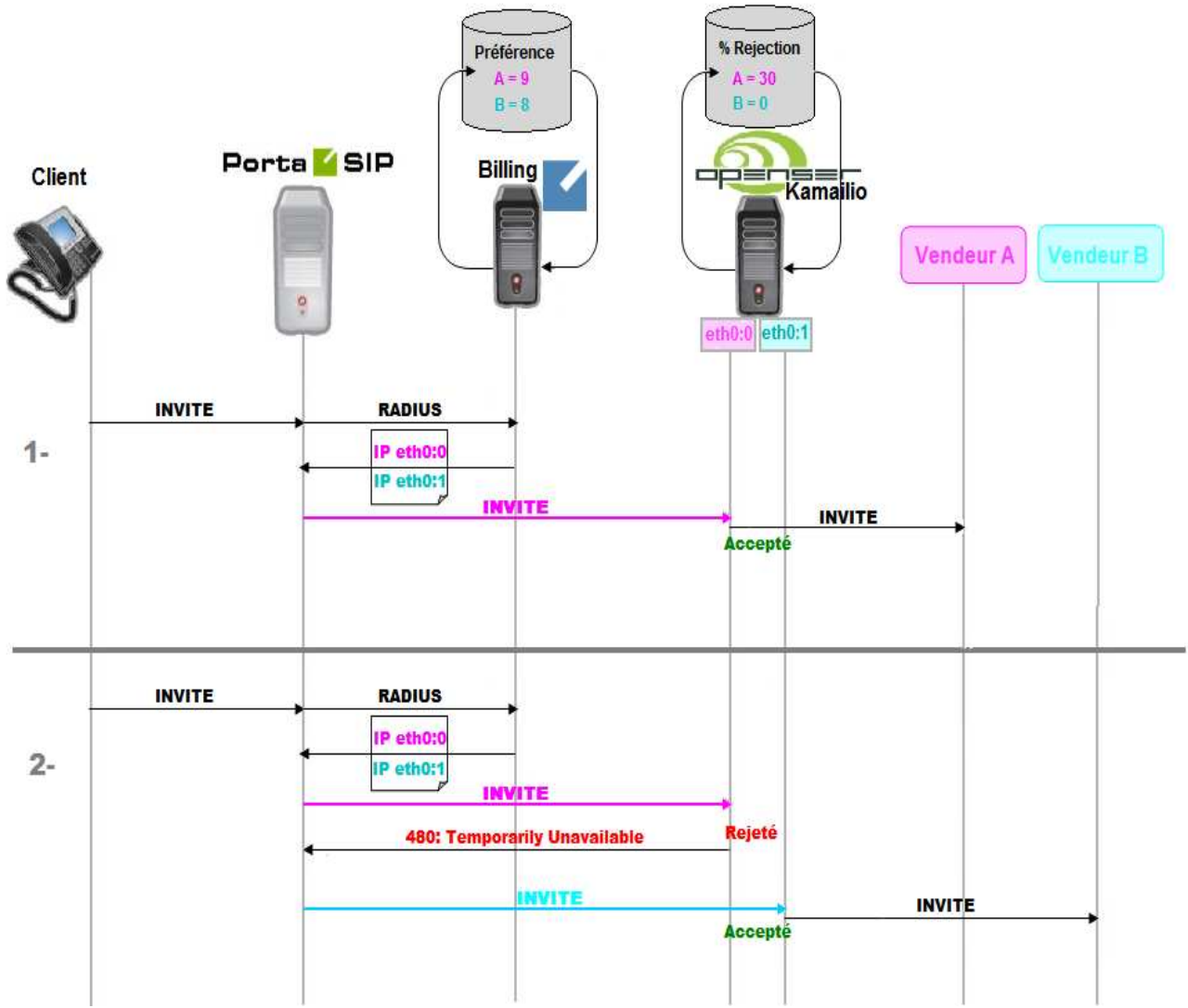

**Fig.8 :** Illustration de deux appels transitant par notre serveur de routage

La probabilité de routage est le complément à 100 du pourcentage de rejet, c'est la probabilité pour que l'appel passe par le vendeur spécifié au premier lieu en cas de rejet la probabilité passe à 100% pour le second vendeur.
L'appel ne sera rejeté qu'une seul fois par notre système de routage.
Le Billing route toujours les appels vers le vendeur dont la préférence est la plus grande mais cette fois le vendeur n'est qu'une interface réseau de notre serveur Kamailio. Ce dernier va décider selon la probabilité de routage spécifié au vendeur caractérisé par l'interface qui vient de recevoir l'appel s'il l'accepte ou non.

• *Principe de fonctionnement:*
Le schéma suivant illustre le scénario du deuxième appel qui a été rejeté la première fois pour qu'il puisse passer au deuxième vendeur dont la préférence est la plus basse.



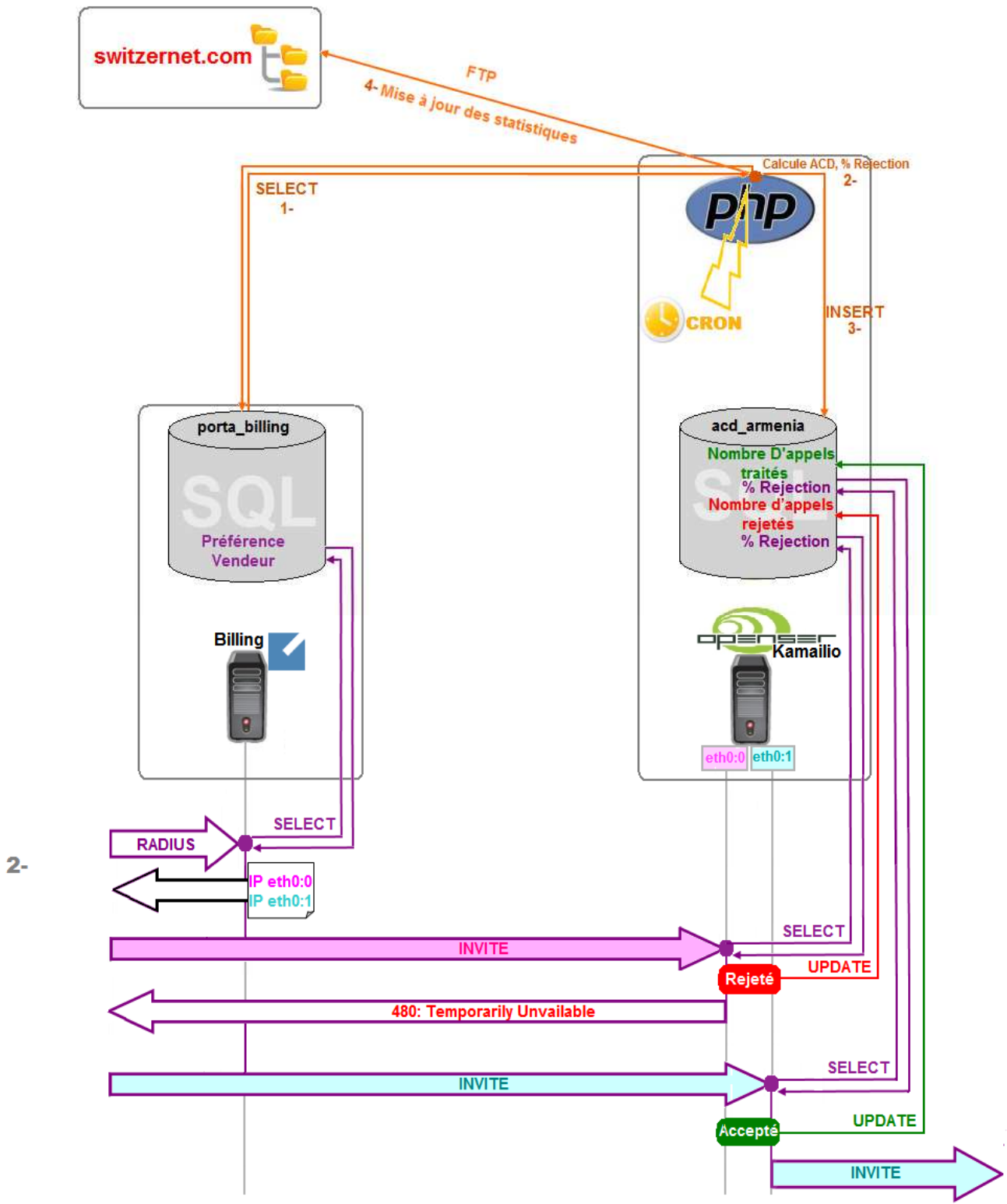

**Fig.9 :** Principe de fonctionnement de routage avec ACD



Un script PHP calcule le pourcentage de rejet et l'ACD pour les X dernières minutes. L'exécution est lancée par CRON chaque 10 minutes mais l'intervalle dont lequel on calcule les variables de décision n'est pas fixe.
La durée minimale de cet intervalle est de 20 minutes et le nombre minimal d'appels est de 20 appels, si l'une de ces deux conditions n'est pas satisfaite alors l'intervalle reste ouvert jusqu'au prochain lancement et c'est pour cette cause qu'on remarque que l'intervalle fait toujours un pas de 10 minutes.
Lorsque ces deux conditions sont vraies l'intervalle est fermé, une nouvelle ligne est ajoutée à la table « acd_vendors » de la base « acd_armenia » avec des nouveaux valeurs qui vont changer le comportement de serveur de routage selon la qualité offerte par les vendeurs pendant ce dernier intervalle.
Enfin, il va uploader un fichier html via FTP aux serveurs web de Switzernet (Switzernet.com et Unappel.ch) qui sert à suivre le comportement du système et pour surveiller son activité ce fichier est mis à jour chaque 5 minutes.
À chaque rejet Kamailio incrémente un compteur présent dans la base de données pour la période en cours et à chaque autorisation il incrémente aussi un autre compteur dans la même base ainsi nous avons le nombre total des appels traités et rejetés par Kamailio.

• ***L'algorithme de calcule du pourcentage de rejection […]***

```
fonction MaxACD (ACD : tableau réel[0..1]) : entier
    début
        si ACD[0] < ACD[1]
            alors retourne 1
            sinon retourne 0
        fin si
    fin
Lexique :
    - ACD : tableau réel[0..1], Moyenne de durée d'appel pour les deux vendeurs.

Algorithme
    Début
        Load_min <- 0.1
        pour i de 0 à 1 faire
            ACD[i] <- ObtenirACD(i)
        Fin pour
        max <- MaxACD(ACD)
        min <- 1 - max
        Rank[max] <- 1
        Rank[min] <- ACD[min] / ACD[max]
        Load[min] <- Load_min + (( 0.5 - Load_min ) * Rank[min] )
        Load[max] <- 1 - Load[min]
        pour i de 0 à 1 faire
            Pref[i] <- ObtenirPref(i)
        Fin pour
        Si Pref[max] > Pref[min]
             alors Rejet[max] <- Load[min] * 100
                   Rejet[min] <- 0
             Sinon Rejet[min] <- load[max] * 100
                   Rejet[max] <- 0
        Fin si
    fin

Lexique :
    - Load_min : réel.
    - ACD : tableau réel [0..1], Moyenne de durée d'appel pour les deux vendeurs.
    - max : entier, ID vendeur avec l'ACD la plus élevé.
    - min : entier, ID vendeur avec l'ACD la moins élevé.
    - Rank : tableau réel [0..1].
```



```
        - Load : tableau réel [0..1].
        - Pref : tableau réel [0..1], Préférence Billing de chaque vendeur.
        - Rejet : tableau entier [0..1], Pourcentage de rejection.
```

Pour démonstration, une calculatrice en ligne est fournie. Le fichier Excel joint montre également les implémentations des équations utilisées pour calculer le taux de rejet en fonction de valeurs de l'ACD. [1]

• *Exemple*

**Fig.10 :** Exemple de calcule du pourcentage de rejection

## 2.3 Développement d'une interface web [5]

Pour afficher les informations des appels stockées dans la base de données « acd_armenia » par
Le script PHP, on a développé une interface web qui permet d'afficher tous les statistiques disponibles déjà uploadé par le script PHP.
Ces statistiques affichent les détails des appels pour chaque intervalle.
Voici les principaux écrans de l'interface web :



## ACD Routing Statistics: Armenia, Yerevan

**November, 2009**

| Mon | Tue | Wed | Thu | Fri | Sat | Sun |
|---|---|---|---|---|---|---|
|  |  |  |  |  |  | 1 |
| 2 | 3 | 4 | 5 | 6 | 7 | 8 |
| 9 | 10 | 11 | 12 | 13 | 14 | 15 |
| 16 | 17 | 18 | 19 | 20 | 21 | 22 |
| 23 | 24 | 25 | 26 | 27 | 28 | 29 |
| 30 |  |  |  |  |  |  |

Select date

**References:**
ACD Routing for Fighting Wrong Signal Supervision [091020 ii]
The software presently running on the transit SIP server [www.kamailio.org]
The billing server applying LCR, preferences, and charging the vendor accounts in this system [www.portaone.com]
Search the number of calls and their duration for a given period [090910]
Voice quality for calls to Armenia [090910]
ASR Verizon for calls to Armenia [090916]
ASR Colt for calls to Armenia [090918]
Billing statistics on the setup time, ASR, and number of concurrent calls [Verizon Colt]

**Fig.11 :** la page d'accueil de statistique

## Traffic load balance with quality routing
[Last update 2009-12-01 00:55:01]

Statistics of calls collected during the previous interval on all vendor connections (passing directly or via ACD routing system). Included are the numbers of calls with zero, problematic, short durations, the conversations, total calls and total minutes of calls ended within (and not during) the previous interval.

Current targets and statistics of calls passing via the ACD routing system during the current interval

| Date and Time | Vendor | Billing priority | =0 | 0< ≤5 | 5< ≤30 | >30 | Calls | Total Minutes | ACD (min) | Target balance | Received during the current interval | Rejected to balance |
|---|---|---|---|---|---|---|---|---|---|---|---|---|
| 2009-11-30 22:10 | Verizon | 9 | 17 | 0 | 1 | 10 | 28 | 142.3 | 12.94 | 56 % | 5 | 2 |
|  | Colt | 8 | 5 | 0 | 2 | 8 | 15 | 109.3 | 10.93 | 44 % |  | 0 |
| 2009-11-30 21:20 | Verizon | 9 | 17 | 0 | 2 | 9 | 28 | 73.5 | 6.68 | 23 % | 21 | 16 |
|  | Colt | 8 | 5 | 0 | 0 | 9 | 14 | 184.1 | 20.46 | 77 % |  | 0 |
| 2009-11-30 20:40 | Verizon | 9 | 18 | 0 | 3 | 10 | 31 | 109.1 | 8.39 | 66 % | 22 | 7 |
|  | Colt | 8 | 7 | 0 | 3 | 4 | 14 | 35.0 | 5.00 | 34 % |  | 0 |
| 2009-11-30 20:20 | Verizon | 9 | 15 | 0 | 0 | 12 | 27 | 130.8 | 10.90 | 43 % | 24 | 14 |
|  | Colt | 8 | 6 | 0 | 2 | 6 | 14 | 104.1 | 13.02 | 57 % |  | 0 |

Bar chart values: 142 / 109; 73 / 184; 109 / 35; 130 / 104

**Fig.12 :** Ecran des détails des appels



• *Résultats* [6]

**Fig.13 :** Vérification des statistiques avec Billing



# Conclusion

La téléphonie sur IP constitue incontestablement une attraction de taille à la fois pour les équipementiers, les opérateurs, les entreprises et le grand public. Si les enjeux économiques justifient largement cette convoitise, il ne faut cependant pas négliger les contraintes techniques à surmonter.

# Référence

**[1]** Rejection Calculator
http://switzernet.com/public/091020-acd-routing/rejection-calc.htm

**[2]** ACD Routing for Fighting Wrong Signal Supervision
http://switzernet.com/public/091020-acd-routing/

**[3]** Le Protocole SIP
http://switzernet.com/company/091023-kamailio/sip/

**[4]**Configuration de Kamailio (OpenSER)
http://switzernet.com/company/091023-kamailio/kamailio/

**[5]**Statistique ACD Routing pour Armenia, Yerevan
http://www.switzernet.com/public/091029-ACDstat/

**[6]**Documentation et téléchargement de Routage Dynamique basé sur la qualité des appels
http://switzernet.com/public/091217-doc-acd-routing/

Routage avec ACD pour Armenie-Yerevan
http://switzernet.com/public/091112-acd-routing/

Converting ACD Routing price list for upload in the billing
http://switzernet.com/company/091117-acd-routing-price-list-process/

Les Testes de ACD-Routage
http://switzernet.com/company/091023-kamailio/kamailio_armenia/

# Glossaire des Acronymes

**A**

**AAA** Authentication Authorization Accounting
**ACD** Average Call Duration

**D**

**DHCP** Dynamic host configuration protocol
**DNS** Domain Name System

**H**

**HTTP** HyperText Transfer Protocol

**I**

**IETF** Internet Engineering Task Force



**IP** Internet Protocol

**M**

**MGCP** Media Gateway Control Protocol
**MMUSIC** Multiparty Multimedia Session Control

**O**

**OSI** Open Systems Interconnection

**P**

**PHP** PHP: Hypertext Preprocessor

**Q**

**QOS** Quality of service

**R**

**RADIUS** Remote Authentication Dial-In User Service
**RFC** Request For Comment
**RTC** Réseau téléphonique commuté
**RTCP** Real-Time Control Protocol
**RTP** Real-Time transport Protocol

**S**

**SCCP** Skinny Client Control Protocol
**SDP** Session Description Protocol
**SER SIP** Express Router
**SIP** Session Initiation Protocol

**T**

**TCP** Transmission Control Protocol
**ToIP** Telephony over Internet Protocol

**U**

**UA** User Agent
**UDP** User Datagram Protocol
**UIT-T** Union Internationale des Télécommunications - normalisation des Télécommunications
**URI** Uniform Resource Identifier

**V**

**VoIP** Voice over Internet Protocol